\documentclass[reprint,amsmath,amssymb,
 aps,pra]{revtex4-1}

\usepackage{graphicx}
\usepackage{epsfig}
\usepackage{color}
\usepackage{amsmath}
\usepackage{txfonts}

\newcommand{\ts}{{\cal Y}}

\begin{document}
\title{The Local Spin Structure of Large Spin Fermions} 
\author{Tin-Lun Ho$^{1,2}$}
\email{ho@mps.ohio-state.edu}
\author{Biao Huang$^1$}
\email{phys.huang.biao@gmail.com}
\affiliation{$^1$Department of Physics, The Ohio State University, Columbus, OH 43210, USA\\
$^2$Institute for Advanced Study, Tsinghua University, Beijing 100084, China}

\date{\today}
\begin{abstract}

We show that  large spin fermions have very rich spin structures. The local spin order of a spin-$f$ Fermi gas is a linear combination of  $2f$ (particle-hole) angular momentum states, $L=1,..,2f$.  $L=1, 2$ represent ferromagnetic and nematic spin order, while $L\geq 3$ are higher spin orders that have no analog in spin-1/2 systems.  Each $L$ spin sector is characterized as $L$ pairs of antipodal points on a sphere.  Model calculations show that some of these spin-orders have the symmetry of Platonic solid, and many of them have non-abelian line defects. 
 \end{abstract}
\maketitle

\section{Introduction}
Prior to the discovery of Bose-Einstein condensates, the only quantum liquids realized experimentally are the electron liquids in solids, and  the low temperature phases of liquid $^{4}$He and $^{3}$He.  All these systems are made up of spin-1/2 fermions (like electrons and and $^{3}$He atoms), or spin-0 bosons ($^{4}$He atoms). Recent advances in the cooling of atomic gases, however, have created the exciting opportunities of studying  high spin  quantum fluids. The spins of  atomic bosons  can range   from $f=1,2$ ($^{87}$Rb bosons) to $f=8$ ($^{162}$Dy bosons)\cite{Dyboson}, and the spin of atomic fermions can be as high as  $f=9/2$ ($^{40}$K) and $f=21/2$ ($^{161}$Dy)\cite{Dyfermion}.  Theoretical generalizations of the Bose-Einstein condensate and fermion pair superfluids to high spin particles have been made about a decade ago \cite{spinorHo, HoYip}. 


While there are many experiments on large spin Bose condensates (or spinor condensates),  experiments on large spin fermions are still at their infancy. At present,  there is no realization of the superfluid phases of large spin fermions because of their very low transition temperatures.  On the other hand, scatterings in different angular momentum channel and dipolar effects can lead to non-trivial spin structures in the normal state, which may be realized at higher temperatures. These scattering lengths in various spin channels can be measured using techniques in ref.\cite{klaus0, klaus1, klaus2}. 
As we shall see, these scattering lengths can lead to a great variety of spin structures in the normal state, most of which do not have analogs in solid state systems. 


The possibility of rich spin structures for high spin fermions has already been illustrated in the cases of spin-$3/2$ fermions \cite{highspintheory1}\cite{highspintheory3}, spin-$9/2$ $^{40}$K experiments\cite{klaus0,klaus1,klaus2}, and alkali earth fermions with SU$(N)$ symmetry\cite{highspintheory2}. The case of spin-3/2 fermions is very illuminating. 
By simply changing the spin value from 1/2 to 3/2,  the system immediately gains a rich  SO$(5)$ symmetry. In this paper, we shall discuss the spin structure of spin-$f$ fermions in the normal state by analyzing their single particle density matrices.  We shall show that these density matrices can be decomposed into different angular momentum components, $L=0,1, ... , 2f$ made up of a particle and a hole. We then show that each $L$-component can be represented by $L$ pairs of antipodal points (or Majorana points) on a sphere.  From the single particle density matrix, one can see that the $L=0$ component is the average density, and the $L>1$ components describe the spin structure of the system.  The entire spin structure is then specified by a sequence of  $2f$ spheres with $1,2,  ..., 2f$ pairs of antipodal Majorana points respectively. To illustrate the special properties of these spin structures, we shall study the class of inert states \cite{inert}  which are robust against perturbations. We show that many of these inert states have the symmetry of Platonic solids and will have non-abelian line defects. 
Furthermore, we shall perform mean field calculations to demonstrate the emergence of some of these inert states.

\section{Symmetry Classifications}
For spin-$f$ bosons, its condensate (known as spinor condensates) is a
$(2f+1)$ component vector  $ \Psi^{}_{m}({\mathbf r})=\langle \hat{\psi}^{}_{m}({\mathbf r}) \rangle$, $m = f, f-1, .. -f$\cite{spinorHo}, where $\hat{\psi}^{}_{m}({\mathbf r})$ is the field operator that destroys a particle with spin component $m$. 
For fermion superfluids made up of pairs of total angular momentum $F$, its  order parameter also transforms like that of a spin-$F$ spinor condensate\cite{HoYip}. 
In this work, we focus on the local spin order of a Fermi  gas, which is contained in the single particle density matrix $\rho_{m_1m_2} = \langle \hat{\rho}_{m_1m_2}\rangle$, 
\begin{equation}\label{orderparameter}
\rho_{m_1m_2}({\bf r}) = \langle  \hat{\psi}^{\dagger}_{m_2} ({\bf r})  \hat{\psi}^{}_{m_1} ({\bf r}) \rangle.
\end{equation} 
For large spin systems where dipolar effects are important, $\rho$ can acquire spatial variations even in equilibrium.  For simplicity, we shall from now on suppress the spatial coordinate until they need to be made explicit. 

Under a spin rotation $\boldsymbol{\theta}$,
the field operator and the density matrix  transform as  
\begin{eqnarray}
\hat{\psi}^{}_{m} & \rightarrow & 
D^{(f)}_{m m'}  (\boldsymbol{\theta})\, \hat{\psi}_{m'}, \\
 \rho_{m_1m_2} &\rightarrow &   D^{(f)}_{m_1m_3}   (\boldsymbol{\theta})\,\,  \rho_{m_3m_4} D^{(f) \dagger}_{m_4m_2}  (\boldsymbol{\theta}),  
\label{rho-trans} 
\end{eqnarray}
where repeated indices are summed over. Here we employ the rotation matrix $D^{(f)}_{mm'}  (\boldsymbol{\theta}) = \langle f m| e^{-i \boldsymbol{\theta} \cdot {\bf F}} |f m'\rangle$ with the hermitian conjugate $D^{(f)\dagger}_{mm'}(\boldsymbol{\theta}) = \langle fm|e^{i\boldsymbol{\theta}\cdot \mathbf{F}}|fm'\rangle$.  ${\bf F}$ is the spin operator for spin-$f$ particles and $|fm\rangle$ is the eigenstate of $(\mathbf{F}^2, F_z)$. To sort out the spin structure of $\rho_{m_1m_2}$, we decompose it in terms of tensor operators of different 
angular momenta (made up of a particle-hole pair). This is achieved by noting that the $(2f+1)\times (2f+1)$ matrix, 
\begin{equation}
 \left( \ts^{(L)}_{M} \right)_{m_1 m_2} \equiv \sqrt{  \frac{2L+1}{2f+1}} \langle f m_1 | L M; f m_2 \rangle, 
\label{Y} \end{equation}
where $ \langle f m_1 | L M; f m_2 \rangle$ is the Clebsch-Gordon coefficient, transforms under rotation as 
\begin{equation}
 D^{(f)}_{m_1m_2} (\boldsymbol{\theta}) \left(\ts^{(L)}_{M}\right)_{m_2m_3} D^{(f)   \dagger}_{m_3m_4} (\boldsymbol{\theta}) = \sum_{M'} \left(\ts^{(L)}_{M'}\right)_{m_1m_4} D^{(L)}_{M' M} (\boldsymbol{\theta}).
\label{Y-trans}  
\end{equation}
Thus, $ (\ts^{(L)}_{M})_{m_1, m_2}$ 
is a tensor operator \cite{CG} (with angular momentum $L$) in the spin-$f$ space. We can then expand $\rho_{m_1m_2}$ as 
\begin{equation}
\rho_{m_1m_2}  = \sum_{L=0}^{2f}\sum_{M=-L}^{L}   \Phi^{(L)}_{M} (\ts^{(L)}_{M}) _{m_1 m_2} .
\label{rho-phi} \end{equation}

To simplify notation, we will sometimes omit the indices $(m_1 m_2)$ for  $(\ts^{(L)}_M)_{m_1m_2}$ and $\rho_{m_1m_2}$, and treat these matrices as $\ts^{(L)}_M$ and $\rho$.
Since the Clebsch-Gordon coefficients are real numbers, so is $(\ts^{(L)}_M)_{m_1m_2}$. That means $\ts^{(L)\dagger}_{M} = \ts^{(L) T }_{M}$, where $\dagger$ and $T$ means hermitian conjugate and transpose of the matrix.  Since the Clebsch-Gordon coefficients satisfies 
\cite{CG}, 
\begin{equation}
\ts^{(L) \dagger}_{M}  
= (-1)^{M}  \ts^{(L)}_{-M},  \,\,\,\,\,\, \,\,\,\, {\rm Tr} \ts^{(L)}_{M} \ts^{(L')\dagger}_{M'} 
=\delta_{L L'} \delta_{MM'}.
\end{equation}
we have 
\begin{equation}
\Phi^{(L)}_{M} = {\rm Tr} \left( \rho \ts^{(L)\dagger}_{M}\right). 
\label{PhiL}
\end{equation}
From the rotational properties in Eqn (\ref{rho-trans}) and (\ref{Y-trans}),  it is easy to show that the vector  
\begin{equation}
\Phi^{(L)}\equiv (\Phi^{(L)}_{-L},\dots ,\Phi^{(L)}_{L} )^T
\end{equation}
transforms as a spin-$L$ vector in spin space, 
\begin{equation}
\Phi^{(L)}_{M} \rightarrow \Phi'^{(L)}_{M} = \sum\limits_{M'}D^{(L)}_{M M'} \Phi^{(L)}_{M'}. 
\label{rot-pro} \end{equation}
Another convenient way to represent this transformation property is to regard $\Phi^{(L)}_{M}$'s as the expansion coefficients of a 
abstract spin state $|\Phi^{(L)}\rangle$ in the angular momentum basis $|L, M\rangle$, 
\begin{equation}\label{sector}
|\Phi^{(L)}\rangle = \sum_{M=-L}^L \Phi^{(L)}_M |LM\rangle.
 \end{equation}
The rotation property of $|\Phi^{(L)}\rangle$ immediately gives Eqn.(\ref{rot-pro}).
Note that not all $\Phi^{(L)}_M$ in $\Phi^{(L)}$ are independent. The fact that $\rho$ is hermitian implies that 
\begin{equation}
\Phi^{(L)\ast }_{M}= (-1)^{M} \Phi^{(L)}_{-M}.
\label{her}
 \end{equation}
{\em This means $\Phi^{(L)}$ is specified by $2L+1$ independent {\em real} variables.  }

Finally, taking the trace of the expansion in Eqn.(\ref{rho-phi}), we have 
\begin{equation}
\Phi^{(0)}=n/(2f+1). 
\label{Phi0} \end{equation}
where we have made use of the fact that 
\begin{equation}
{\rm Tr} \ts^{(L)}_{M}=0, \,\,\,\,  {\rm for}  \,\,\,\, L\geq 1.
\label{YL}\end{equation}
To show Eq.(\ref{YL}), we  take the Trace of Eqn.(\ref{Y-trans}). This gives  ${\rm Tr}\ts^{(L)}_{M} = {\rm Tr}\ts^{(L)}_{M'} D^{(L)}_{M' M} (\boldsymbol{\theta})$ for all $L\geq 1$ and all $\boldsymbol{\theta}$, which can only be satisfied if Eqn.(\ref{YL}) is valid. 
Because of Eqn.(\ref{Phi0}), one sees that the non-trivial spin structures are given by the traceless part of $\rho$, 
\begin{equation}\label{nontrivialorder}
\tilde{\rho}_{mn} \equiv \rho_{mn} - \frac{n}{2f+1} \delta_{mn}. 
\end{equation}


Although $\Phi^{(L)}$ is formally similar to a spin-$L$ spinor condensate, it has very different meaning. 
From Wigner-Eckart theorem,  we note that  ${\cal Y}^{(L)}_{M}$ is proportional to 
 a product of $L$ spin operators ${\bf F}$ in spin-$f$ space. Thus, 
$\Phi^{(1)}$ and $\Phi^{(2)}$ are proportional to a single and two ${\bf F}$ operators respectively, and thus
represent  ferromagnetic and nematic order respectively in spin-$f$ space. 
The vectors  $\Phi^{(L)}$  for $L\geq 1$ will be referred to as  the $L$-th spin order of the system, $L=1,2, ..$, which are all contained in the traceless part of $\tilde{\rho}$.

To gain further insight, we express each $\Phi^{(L)}$ in Majorana representation as a set of  $2L$ points (referred to as Majorana points) on the unit sphere $S_2$ \cite{Majorana, Demler, Ueda}. To accomplish it, we use the  Schwinger bosons representation of 
 angular momentum states, 
 \begin{equation}
 |LM \rangle   = \frac{a^{\dagger L+M} b^{\dagger L-M}  }{\sqrt{ (L+M)! (L-M)!} } |0\rangle,   \hspace{0.3in}
 \end{equation}
 where $a$ and $b$ are boson operators, and $(a,b)^{T}$ transform as a spin-1/2 spinor  \cite{sakurai}. 
A general state of the form Eqn.(\ref{sector}) can then be factorized as 
\begin{equation}
|\Phi^{(L)} \rangle =
 \lambda^{(L)} \prod^{2L}_{i =1} ( u^{(L)}_{i}  a^{\dagger} + v^{(L)}_{i}  b^{\dagger})|0\rangle,    \label{Sch} 
 \end{equation}
where $\lambda^{(L)}$ is a constant, and $\zeta_{i}^{(L)} \equiv  (u^{(L)}_{i}, v_{i}^{(L)})^{T}$ is a normalized spinor
 \begin{equation}
 |u^{(L)}_{i}|^2 + |v^{(L)}_{i}|^2 =1. 
 \end{equation}
 Eqn.(\ref{Sch}) follows from the fundamental theorem of algebra which implies a homogenous polynomial ${\cal P}(x,y) = \sum_{M=-L}^{L} \alpha_{M} x^{ L+M}_{} y^{L-M}$ 
 can be factorized to $2L$ linear terms ${\cal  P}(x,y) = \lambda \prod_{i=1}^{2L}(u_{i}x + v_{i}y)$. 

To simplify notations, we shall suppress the superscript $^{(L)}$ when we discuss a specific $L$ component. It will be reinstated when needed. Using the standard representation for a spinor 
\begin{equation}
\zeta= (u, v)^{T}\equiv ({\rm cos}\frac{\theta}{2} e^{-i\phi/2}, {\rm sin}\frac{\theta}{2} e^{i\phi/2} )^T e^{i \chi/2},  
\end{equation}
we have 
\begin{equation} 
\zeta^{\dagger}\vec{\sigma}\zeta = \hat{\bf n} = {\rm cos}\theta \hat{\bf z} + {\rm sin}\theta ({\rm cos}\phi \hat{\bf x} 
+ {\rm sin}\phi\hat{\bf y}). 
\end{equation}
Hence each $\zeta_{i}$ in Eqn.(\ref{Sch}) can be represented as a point on the unit sphere $S_2$ in the direction $\hat{\bf n}_{i}$ with polar angle  $(\theta_{i}, \phi_{i})$. Note that all  the phases $\chi_{i}$ are absorbed into the constant $\lambda$. 

However, not all $\hat{\bf n}_{i}$'s are independent, as
Eqn.(\ref{her})  implies that Eqn.(\ref{Sch}) can be rewritten as 
 \begin{eqnarray}\nonumber
|\Phi^{(L)}\rangle &=  \sum\limits_{M=-L}^{L} \frac{\Phi^{(L)\ast}_{M} }{\sqrt{ (L+M)! (L-M)!} }   b^{\dagger L+M} (-a^{\dagger})^{L-M} |0\rangle     \nonumber \\
 &= \lambda^{\ast}  \prod^{2L}_{i =1} ( u_{i}^{\ast}  b^{\dagger} - v_{i}^{ \ast}  a^{\dagger})|0\rangle
  \hspace{0.9in}
\label{uv-sym} 
\end{eqnarray} 
where we have suppressed the superscript $^{(L)}$. 
Eqn.(\ref{uv-sym}) shows that the spinors $\zeta_i$ in Eqn.(\ref{Sch})  must be accompanied by its time reversed partner $-i\sigma_{2}\zeta_i^{\ast}= (-v^{\ast}_{i}, u^{\ast}_{i})$.
 Therefore, the $2L$ vectors $\hat{\bf n}_{i}$  must appear in antipodal pairs  $(\hat{\bf n}_{i}, -\hat{\bf n}_{i})\equiv [\hat{\bf n}_{i}]$. 
 It is then sufficient to represent each pair by only one of its members.  The presence of antipodal pairs implies 
Eqn.(\ref{Sch}) is of the form
\begin{eqnarray} \nonumber
|\Phi^{(L)} \rangle &=& 
\lambda^{(L)} \prod^{L}_{i =1} ( -u^{}_{i} v^{\ast }_{i} a^{\dagger 2} + (|u_{i}|^2 - |v_{i}|^2) a^{\dagger}b^{\dagger} + u^{\ast}_{i} v^{}_{i} b^{\dagger 2} )|0\rangle
\\ \nonumber
&=&
\frac{\lambda^{(L)}}{2} \prod_{i=1}^L \left(\sin\theta_i e^{i\phi_i} b^{\dagger 2} \right.
\left. - \sin\theta_i e^{-i\phi_i} a^{\dagger 2} + \cos_i\theta_i a^{\dagger}b^{\dagger}    \right) |0\rangle
\\  \label{new}
 \end{eqnarray}
with $\lambda^{(L)\ast }=\lambda^{(L)}$ because of Eqn.(\ref{her}). 
 The $(2L+1)$ real variables of $\Phi^{(L)}$ is now represented by the $L$ unit vectors $\hat{\bf n}_{i}$ and a real number $\lambda^{(L)}$. 
In equilibrium,  different $\lambda^{(L)}$'s are correlated  to minimize the energy.



In summary, we have decomposed the non-trivial traceless part of the density matrix $\tilde{\rho}_{m_1m_2}(\mathbf{r})$, which represent local spin order into various spin-L vectors $\Phi^{(L)}$, where $L=1,2,\dots 2f$. Each $\Phi^{(L)}$ is represented by $L$ antipodal pairs of points on a spherical surface, whose radius $\lambda^{(L)}$ represents the strength of the L-sector of spin order. In the following, we discuss the properties of the local spin order in each $L$-sector. 
 

\section{Topological defects and spin texture of different spin order sectors}
 For $L=1$, the ferromagnetic order, 
there is only one pair of Majorana points $[\hat{\bf n}]$.  (See Fig. 1(i)). Since $\hat{\bf n}$ can be in any direction, the configurational space is the unit sphere $S_2$. Note that $[\hat{\bf n}]$ and $[-\hat{\bf n}]$ are distinct because 
$|\Phi^{(1)}\rangle$
 becomes $-|\Phi^{(1)}\rangle$ as 
$\hat{\bf n}$ changes continuously  to $-\hat{\bf n}$ \cite{L=1}. Since the first homotopy group of $S_{2}$ is trivial, 
($\pi_1(S_{2})=0$), the vector field $\{ \Phi^{(1)}_{M}({\bf r}) \}$ has no topologically stable line defects\cite{Mermin}. 

For $L=2$, the nematic order, there are two Majorana pairs $[\hat{\bf n}] [\hat{\bf m}]$. 
If  $\hat{\bf n} = \pm \hat{\bf m}$, the system is  uniaxial nematics characterized by  a single antipodal pair on the unit sphere
with each pole doubly occupied. (See Fig.1(ii)). 
Unlike $L=1$, where $[\hat{\bf n}]$ and  $[-\hat{\bf n}]$ are distinct, the states $[\hat{\bf n}][\hat{\bf n}]$  and $[-\hat{\bf n}][-\hat{\bf n}]$  are identical, as they correspond to the same state 
$|\Phi^{(2)}\rangle = \lambda (ua^{\dagger}+vb^{\dagger})^2(-v^{\ast}a^{\dagger}+ u^{\ast}b^{\dagger})^2|0\rangle$.  (Note that ($[\hat{\bf n}][-\hat{\bf n}]= - [\hat{\bf n}][\hat{\bf n}]$). The configuration space is therefore $S_2$ with antipodal points identified, which is the projected space $P_2$. Since $\pi_1(P_2)=Z_2$, there is only one type of nontrivial line defect. 
If $\hat{\bf n} \neq \pm \hat{\bf m}$, it is straightforward to show that $|\Phi^{(2)}\rangle$ is unchanged only under a $\pi$ rotation along the orthogonal axes $\hat{\bf n}\times \hat{\bf m}$, $\hat{\bf n}+\hat{\bf m}$, and $\hat{\bf n}-\hat{\bf m}$.  (See \cite{L=2} and see Fig. 1(iii)). The system is therefore a bi-axial nematics, and has nonabelian line defects\cite{Mermin}. 

For $L\ge 3$, there will be more pairs of Majorana points. A simple case is that all pairs locate at the same position, as what happens for uniaxial nematics. The discussions above  show that for odd and even $L$,  the  configuration space  is $S_{2}$ and $P_2$ respectively. In general, the pairs of points can distribute arbitrarily, forming the vertices of an irregular polygon, as those in Fig. 1 (e).  The fact that {\em the Majorana points must appear in antipodal pairs forbids the polygon to have tetrahedral symmetry}, as shown in (Fig.1 (iv)). This is different from the situation in bosonic spinor condensates where tetrahedral symmetry is allowed in the case when spin $S\geq 2$\cite{Demler}.

Of particular interests are the cases when the Majorana points are distributed in high symmetry, such as the Platonic solids shown in Fig. (c) (d) (f) (g), which are cube, octahedron, isocahedron, and dodecahedron respectively. The symmetry groups of (c) and (d) is the  ocatahedral group $O$, and that for (f) and (g)  is isocahedral group $Y$. These states belong to the class of ``inert states'' whose structures (i.e. distribution of Majorana points) are independent of interaction parameters\cite{inertstates}. These states, if present, must therefore occupy a finite region in parameter space, and have a good chance of being observed. 
We show in the next section through a mean field calculation that all these Platonic solids states can arise from spin exchange interaction\cite{spinorHo}. 


 
As mentioned before, the spin order is specified by the set of vectors $\{ \Phi^{(L)}\}$ with $L=1, 2, .. 2f$. In general, when dipolar interaction is taken into account, these vectors (and their corresponding Majorana points in $S_{2}$) will vary in space, forming a spin texture in each $L$-sector.  
The general behavior of these spin textures $\{ \Phi^{(L)}({\bf r}) \}$ is illustrated in  Figure 1 (1-7) for the case of  $f=7/2$. 
The figure displays the  spin orders $|\Phi^{(L)}({x})\rangle$ along a loop ${\cal C}$ in real space, which can be represented as a straight line along $x$ with end points identified. The entire set of  spin order  $\{ \Phi^{(L)}(x), L=1,2, ..2f \}$  is  represented as an array of $2f$  spherical surfaces with radius $|\lambda^{(L)}(x)|$ and $L$ pairs of antipodal Majorana points. One can recall that the spin texture in spin-1/2 systems corresponds to a rotation of a vector in space; the direction of the vector designates the local spin order $\langle \mathbf{S} \rangle$. In comparison, the spin textures in higher $L$-sector of the spin order correspond to the rotation (or even deformation) of a polygon in space.
As long as $\lambda^{(L)}(x)\neq 0$ and different Majorana pairs do not merge as on traverses the loop ${\cal C}$, each $L$ sector can have its own line defects.

\begin{widetext}

\begin{figure}[t]
\includegraphics[width=16cm]{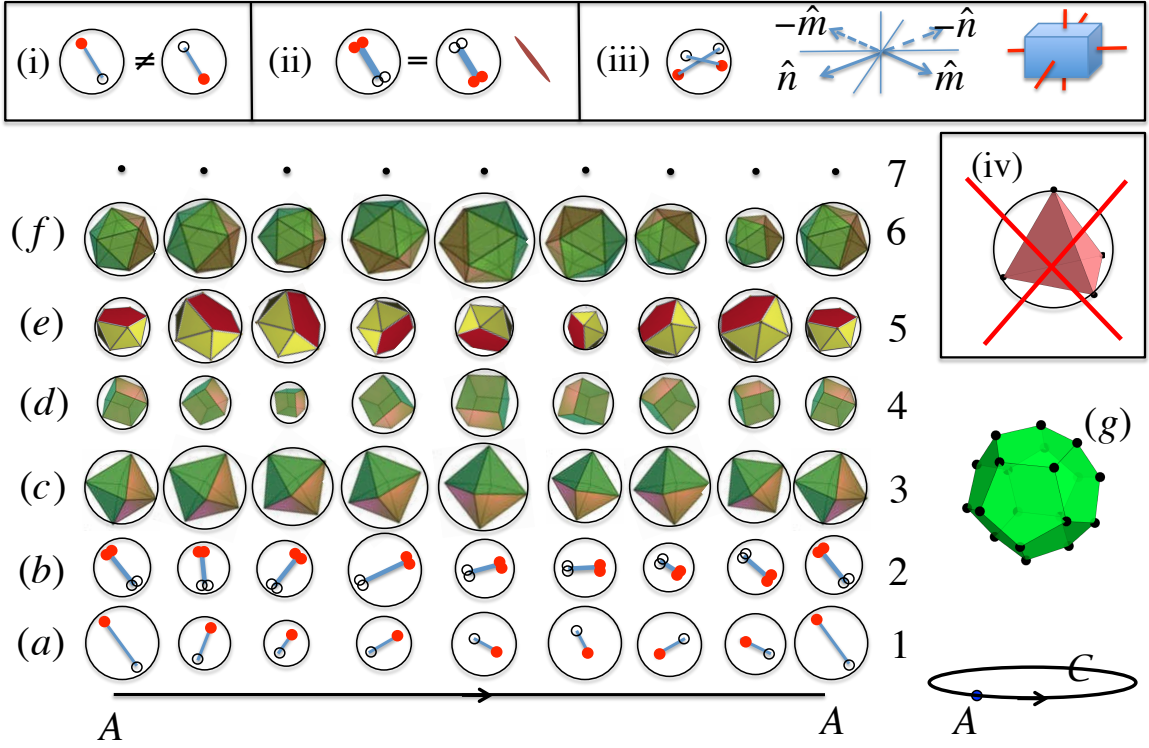}
\caption{   Inset (i): Ferromagnetic order $\Phi^{(1)}$  represented the Majorana pair $[\hat{\bf n}]\equiv(\hat{\bf n}, -\hat{\bf n})$. $[\hat{\bf n}]$ and $[-\hat{\bf n}]$ are distinct. Inset (ii) : Uniaxial nematic state $\Phi^{(2)}$  represented by two identical Majorana pairs:  
It has the symmetry of a rod.  $[\hat{\bf n}][\hat{\bf n}]$ and $[-\hat{\bf n}][-\hat{\bf n}]$ are identical. 
(iii) Biaxial nematic state $\Phi^{(2)}$ with two distinct Majorana pairs:  The quantum state is invariant under $\pi$ rotation about 
$\hat{\bf n}\times \hat{\bf m}$, $\hat{\bf n}+\hat{\bf m}$,  $\hat{\bf n}-\hat{\bf m}$. It has the symmetry of a brick with different edge lengths. 
Inset (iv): The Majorana points can not form a tetrahedron because its vertices do not form antipodal pairs. 
The numerals $(1,2, ..7)$ denote the spin order $(\Phi^{(1)}, \Phi^{(2)}, ... \Phi^{(7)})$of a spin $f=7/2$ Fermi gas along the loop ${\cal C}$ in real space. 
$\Phi^{(L)}_{M}({\bf R})$ is represented by a sphere $S^{(L)}$ of radius $|\lambda^{(L)}({\bf R})|$ marked with $L$ pairs of antipodal points.  Here, $\lambda^{(7)}=0$,  
$\Phi^{(3)}$,  $\Phi^{(4)}$,  $\Phi^{(6)}$ form an octahedron, a cubic, and  an icosahedron. 
$\Phi^{(5)}$ is a polyhedron with 5 vertices forming a pentagon. 
The texture of $\Phi^{(2)}_{M}(x)$ depicted implies a line defect inside loop {\cal C}, whereas the texture of $\Phi^{(1)}_{M}(x)$ is defect free. 
Our  model calculations for spin $f=21/2$ Fermi gas reveal Platonic solid  configurations like (c), (d), (f) and  dodecahedron (g) in certain parameter regimes. 
} \label{fig1}
\end{figure}
\end{widetext}

\section{Energetic considerations and Platonic solid inert states} 
Next, we discuss how interaction effects give rise to the spin order discussed above. We shall consider a general short range spin-exchange interaction between fermions\cite{spinorHo}.  Such a description has been shown to be effective in recent experiments in $^{40}$K\cite{klaus0,klaus1,klaus2}, where dipolar interactions are negligible. Many other high spin systems (i.e.$^{161}$Dy) have stronger dipolar interactions which will lead to non-uniform spin ordering. As a first step, we shall ignore dipolar interactions. In practice, dipolar interactions can be averaged out to zero through a sequence of magnetic pulses\cite{dipole}. On the other hand, many competing orders may arise at low temperatures, including the superconducting ordering. Different orders would be favored in different regions in the parameter space. For instance, for spin-1/2 fermions, attractive interaction leads to superconducting phase while repulsive interaction induces magnetic ordering \cite{attractiverepulsive}. For higher spin systems, in general, the orders may have overlap in the parameter space and different orders will compete with each other. Here we do not consider the possibility of competing phases, and discuss first the parameter regions that can give rise to spin ordering.

\subsection{ Mean Field Phase Diagram}
The Hamiltonian for local spin-exchange interaction is $H = H_0 + H_1$, where the kinetic and interaction parts are
\begin{eqnarray}
H_0 &=& \int d^3 \mathbf{r} \sum\limits_m \psi_m^\dagger(\mathbf{r}) \left( \frac{\hbar^2\nabla^2}{2\tilde m} - \mu\right) \psi_m(\mathbf{r}) \\ \label{interactionh}
H_1 &=&
\int d^3 \mathbf{r}
 \sum\limits_{\substack{m_1m_2\\m_3m_4}}  
 \psi_{m_1}^\dagger (\mathbf{r}) \psi_{m_2}^\dagger(\mathbf{r}) \gamma_{m_1m_2 m_3m_4} \psi_{m_4}(\mathbf{r}) \psi_{m_3}(\mathbf{r}),\qquad
\end{eqnarray}
where $\tilde{m}$ is the fermion mass, $\mu$ is the chemical potential, and 
\begin{equation}
\gamma_{m_1m_2m_3m_4} = \frac{1}{2} \sum\limits_{F=0,2,\dots}^{2f-1} g_F \sum\limits_{M_F=-F}^F \langle m_1m_2|FM_F\rangle\langle FM_F|m_3m_4\rangle.
\end{equation}
Here $\langle m_1 m_2|FM_F\rangle$ is an abbreviation for Clebsch-Gordon coefficients $\langle fm_1;fm_2|FM_F\rangle$. $g_{F}\equiv 4\pi \hbar^2 a_{F}/\tilde{m}$ is the interaction constant in the scattering channel with total spin-$F$, and $a_{F}$ is the corresponding scattering length. 
For half integer spin $f$, $\langle fm_1; fm_2|FM_F\rangle = \pm \langle fm_2;fm_1|FM_F\rangle$ for odd/even $F$ respectively. Thus, (\ref{interactionh}) and Fermi statistics require $F$ to be even integers only. That means $\gamma_{m_1m_2m_3m_4}$ is odd under the exchange $m_1\leftrightarrow m_2$ or $m_3\leftrightarrow m_4$.

\begin{widetext}

\begin{table}
\caption{Exploration of $T=0$ mean field phase diagram for spin $f=21/2$ fermion systems in the presence of spin exchange interactions. $k_n = (6\pi^2 n)^{1/3}$, where $n$ is the {\em total} number of particles in the system. $\{ a_F , F=0,2,\dots,2f-1 \}$ are scattering channels with total spin $F$. The spin orders of different sectors are represented in Majorana representation. $|\Phi^{(L)}\rangle$ sector would contain $2L$ points ($L$ pairs). But different points may occupy  the same location, as is the case in (I) and (II). The objects in set (II) and (III) are Platonic solids. All of these states are inert states.}
\label{tab:phasediagram}
\begin{tabular}{l | c c c c c c c c c c c | c}
\hline \hline
Set & $k_n a_0$ & $k_n a_2$ &$k_n a_4$ & $k_n a_6$ &$k_n a_8$ & $k_n a_{10}$ &$k_n a_{12}$ & $k_n a_{14}$ &$k_n a_{16}$ & $k_n a_{18}$ &$k_n a_{20}$ & Sectors of Spin Order\\
(I) & 0.1 & 0.3 & 0.4 & 0.45 & 0.5 & 0.5 & 0.45 & 0.45 & -0.3 & -0.5 & -0.5 & 
\raisebox{-0.25cm}{\includegraphics[width=1cm]{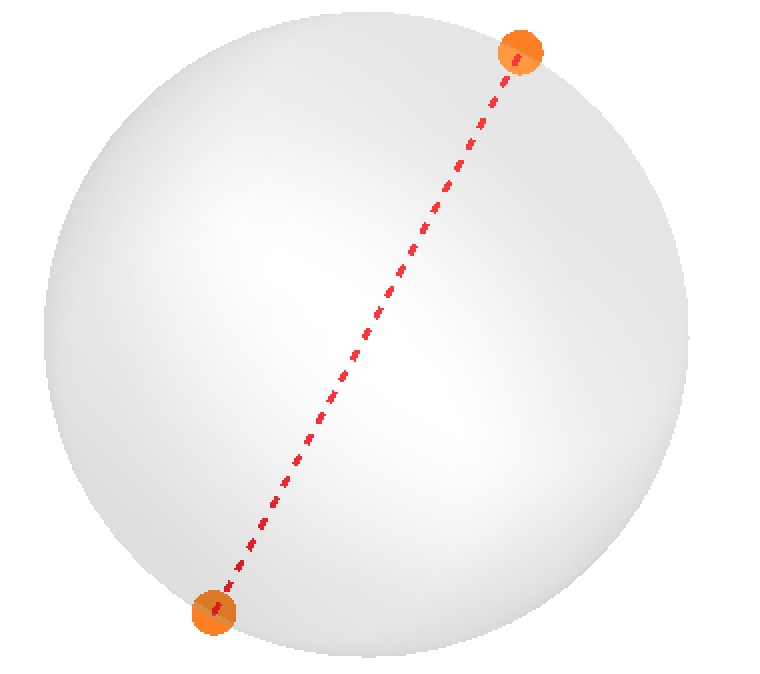} } $|\Phi^{(1)}\rangle, |\Phi^{(2)}\rangle, |\Phi^{(3)}\rangle, |\Phi^{(4)}\rangle$\\
(II) & -0.1 & -0.6 & -0.6  &  0.55 &  0.75 &  0.75 & 0.45 &  -0.75 &  -0.8 & 0.8 & 0.8 & 
\raisebox{-0.25cm}{\includegraphics[width=1cm]{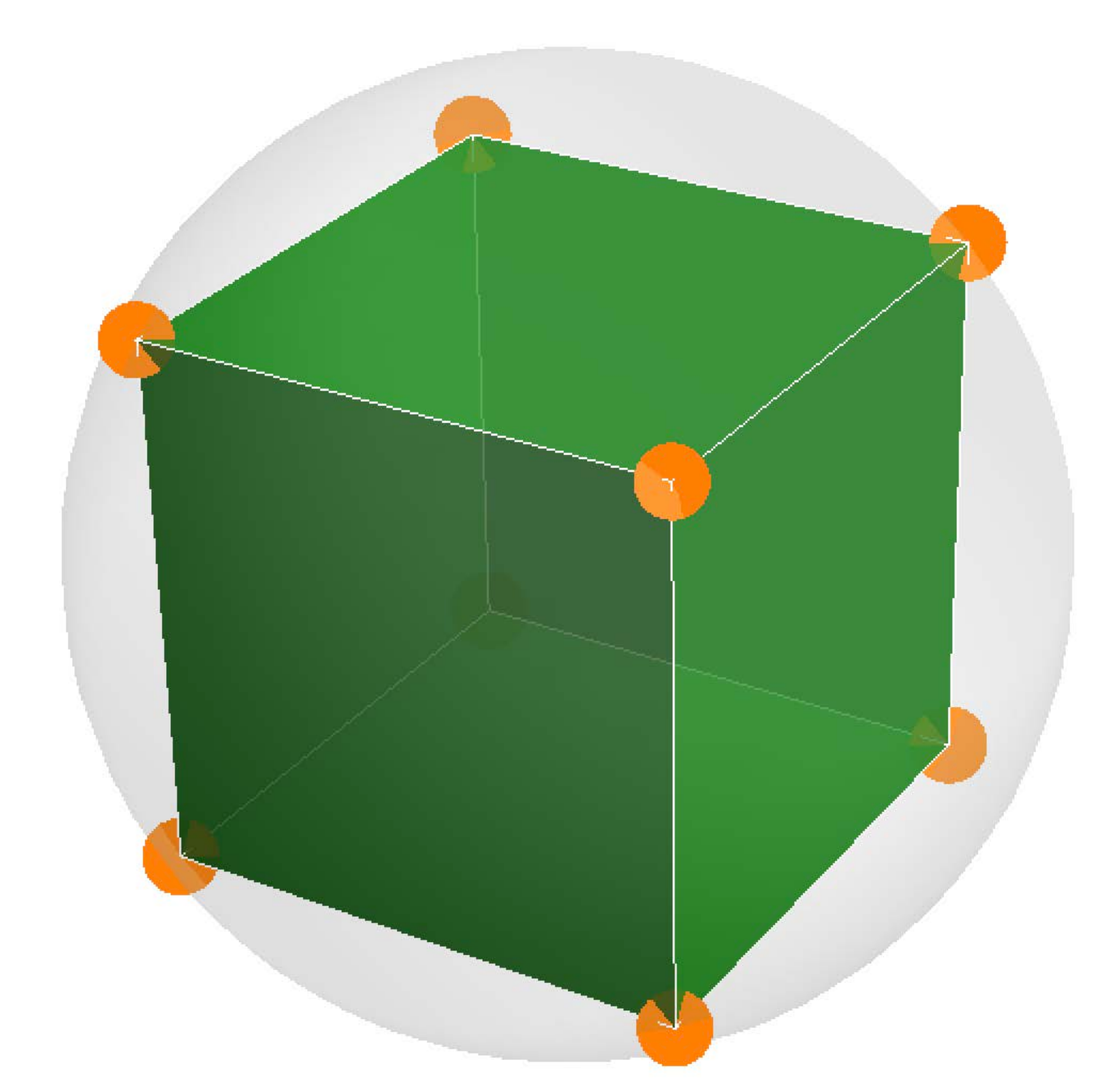} } $|\Phi^{(4)}\rangle, |\Phi^{(8)}\rangle$ and 
\raisebox{-0.25cm}{\includegraphics[width=1cm] {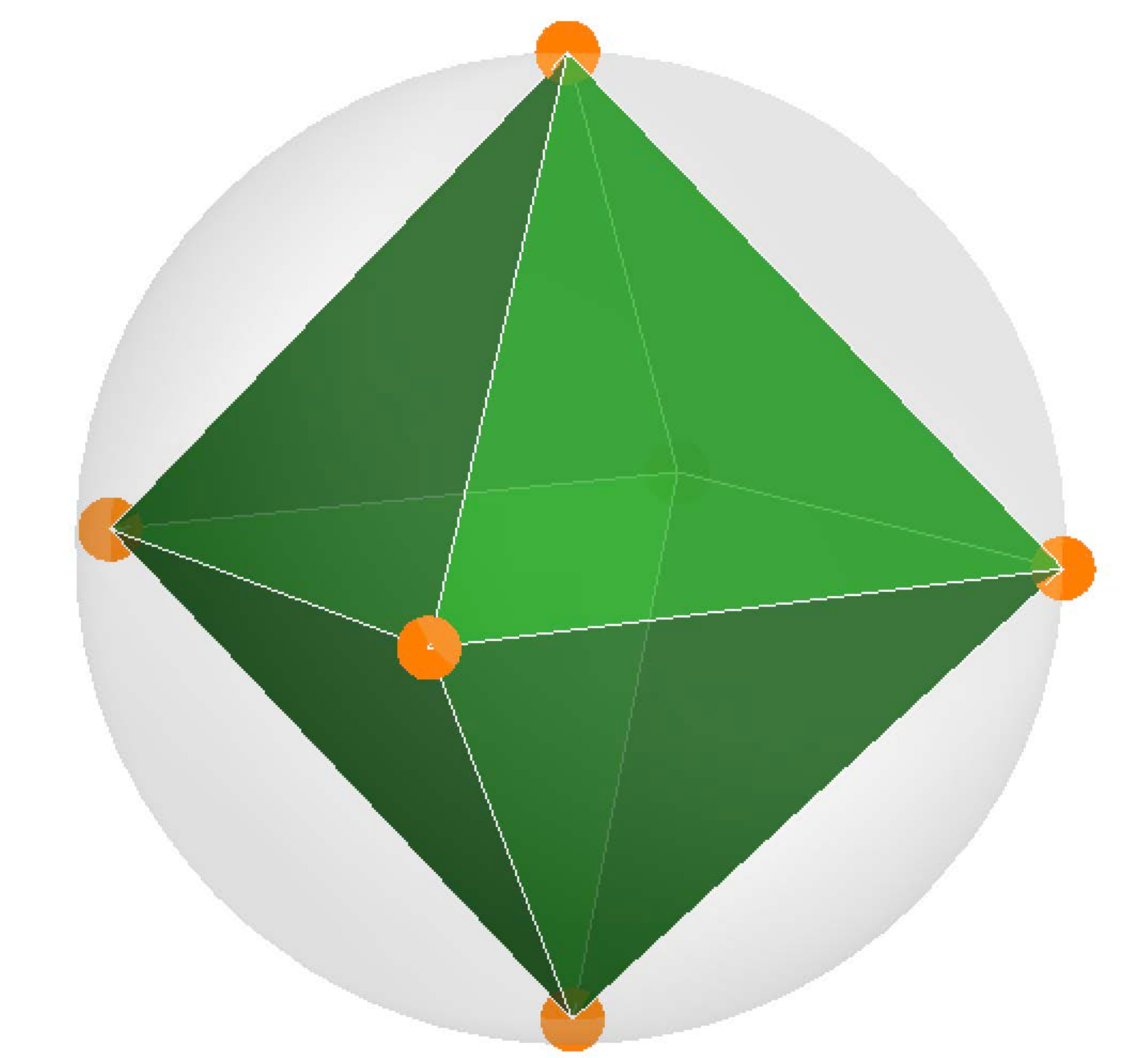} } $|\Phi^{(6)}\rangle$\\
(III) & -0.6 &  -0.74 &  0.87 & 0.79 & 0.84 & -0.8 & -0.83 & 0.82 & 0.82 &  -0.85 & 0.85 &
\raisebox{-0.25cm}{\includegraphics[width=1cm]{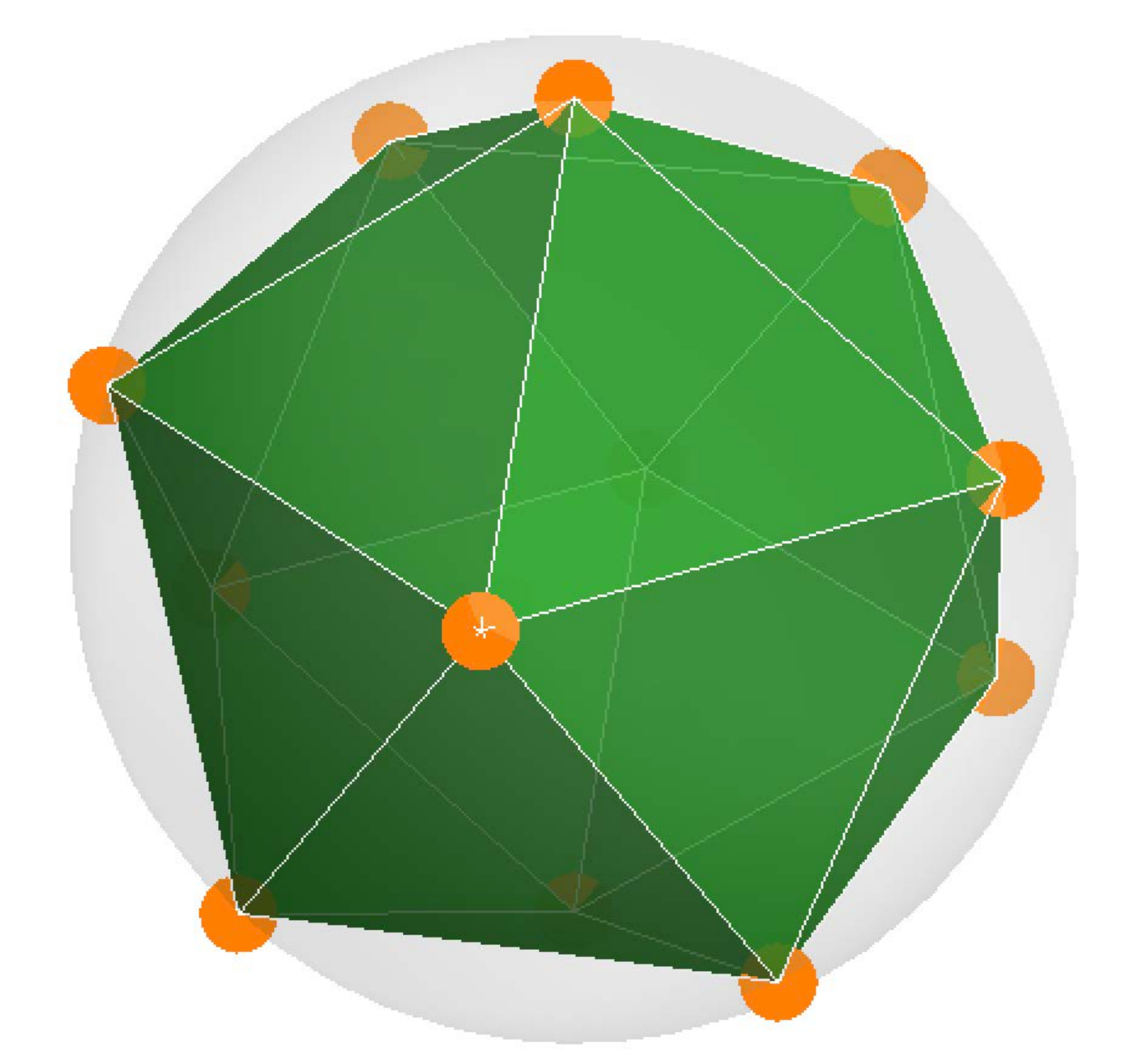} } $|\Phi^{(6)}\rangle$ and 
\raisebox{-0.25cm}{\includegraphics[width=1cm]{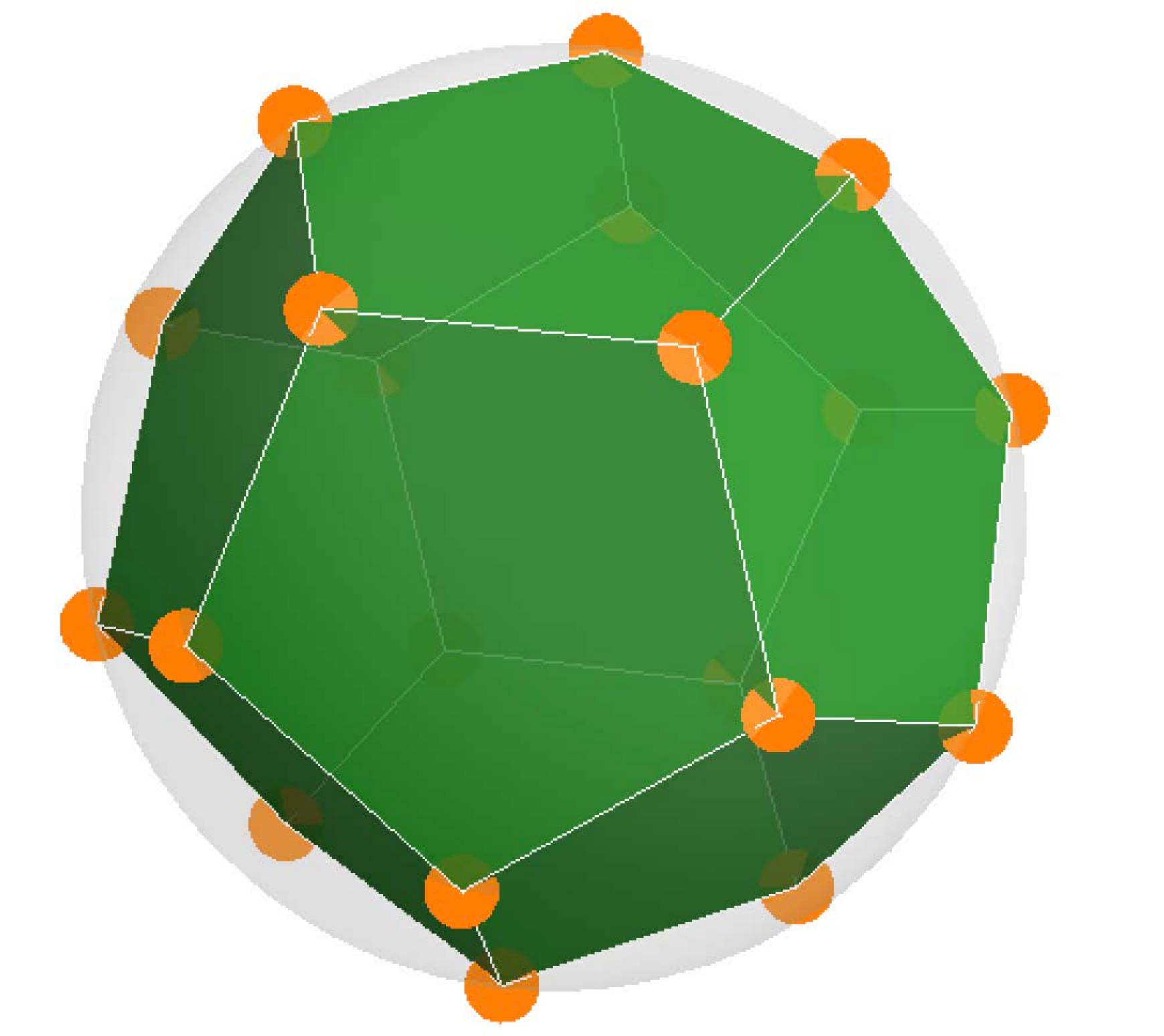} } $|\Phi^{(10)}\rangle$\\
\hline
\end{tabular}
\end{table}

\end{widetext}

(It is useful here to summarize our notations. $(f,m)$ are the eigenvalues of single atom spin operators $(\mathbf{F}^2, F_z)$. $(F, M_F)$ is the total spin and magnetic quantum number when two atoms scatter with each other. $(L,M)$ introduced previously denote the particle-hole total angular momentum quantum numbers, and is the quantum numbers we use to classify spin orders in terms of different sectors $|\Phi^{(L)}\rangle$ in (\ref{sector}). )

We shall study the {\em uniform} spin order using the  mean field approximation.  The order parameter is $\rho_{m_1m_2}$ defined in (\ref{orderparameter}), and it is equal to the average
\begin{equation}
\rho_{m_1m_2} =  \int \frac{d^3\mathbf{r}}{V}  \,  \rho_{m_1m_2} (\mathbf{r}) = \frac{1}{V}\sum\limits_\mathbf{k} \left\langle c_{\mathbf{k} m_2}^\dagger c_{\mathbf{k} m_1}\right\rangle,
\end{equation}
where we used the Fourier tranform $\psi_{m}(\mathbf{r}) = \frac{1}{\sqrt V} \sum\limits_\mathbf{k}e^{i\mathbf{k} \cdot \mathbf{r}}c_{\mathbf{k}m}$, and $V$ is the volume. 
Then the mean field Hamiltonian reads
$H_{\mbox{\tiny MF}} =
\sum_{\mathbf{k}, m_1 m_2} c_{\mathbf{k} m_1}^\dagger  \mathcal{H}_{m_1m_2}(\mathbf{k}  ) c_{\mathbf{k} m_2}$, with 
\begin{eqnarray}\label{mfh}
\mathcal{H}_{m_1m_2} (\mathbf{k}) &=& 
 \left(\varepsilon_\mathbf{k} -\mu \right)\delta_{m_1m_2} +4\Gamma_{m_1m_2}
 ,  \\
\Gamma_{m_1m_2} &=& \sum_{m_3m_4} \gamma_{m_1m_3m_2m_4} \rho_{m_4m_3}. 
\end{eqnarray}
Here $\varepsilon_{\mathbf{k}} = \hbar^2 k^2/2\tilde{m}$ is the kinetic energy. The quardratic Hamiltonian can be diagonalized in spin space through a unitary transform $(U^\dagger \mathcal{H(\mathbf{k}}) U)_{m_1m_2} =(\epsilon_\mathbf{k} -\mu_{m_1}[\Gamma]) \delta_{m_1m_2}$, where $\mu_{m}[\Gamma] = \mu - 4(U^\dagger\Gamma U)_{mm}$. Then the quasi-particles $b_{\mathbf{k}m_1} =\sum_{m_2} U^\dagger_{m_1m_2}c_{\mathbf{k}m_2}$ are free fermions obeying
$\frac{1}{V}\sum_\mathbf{k} \left\langle b_{\mathbf{k}m_1}^\dagger b_{\mathbf{k}m_2}\right\rangle = n_{m_1}\delta_{m_1m_2}$, where $n_{m_1} = \frac{1}{V}\sum_\mathbf{k}\left( e^{(\epsilon_\mathbf{k}-\mu_{m_1}[\Gamma])/k_BT}+1\right)^{-1}$.
With the above information, we can determine the order parameter through the consistency equation
\begin{equation}\label{consistency}
\rho_{m_1m_2} = \sum\limits_{m_3} U_{m_1m_3} n_{m_3} U^\dagger_{m_3m_2}, 
\end{equation}
and obtain the spinor vectors $\{ \Phi^{(L)} \}$ using Eqn.(\ref{PhiL}). 

We have solved the self consistency equation equation (\ref{consistency}) at $T=0$ numerically for the case of $f=21/2$ with some specific value of gas parameters $\{ k_{n} a_{F} \}$'s, where
\begin{equation}
k_{n} = (6\pi^2 n)^{1/3},
\end{equation} 
and $n$ is the total number density.  See Table \ref{tab:phasediagram}.
Since $a_{F}$'s are unknown at present, we have tried various parameter sets, labelled (I) to (III) in Table \ref{tab:phasediagram}. Their mean field states are:

\noindent (I):  Only $\Phi^{(1)}, \Phi^{(2)}, \Phi^{(3)}, \Phi^{(4)}$ are nonzero.  The Majorana points of each one of them collapse into a single antipodal  pair
like (a) and (b) in Fig. 1. The pairs  of different $L$  orient differently. 

\noindent  (II): Only $\Phi^{(4)}, \Phi^{(6)}$, $\Phi^{(8)}$ are non-zero.  $\Phi^{(4)}$ and $\Phi^{(8)}$ form cubes  (Fig.1(d)). $\Phi^{(6)}$ forms an octahedron (Fig.1(c)). For $\Phi^{(8)}$  and $\Phi^{(6)}$, the vertices of the cube and octahedron are doubly occupied respectively.  

\noindent  (III): Only $\Phi^{(6)}$ and $\Phi^{(10)}$ are non-zero. $\Phi^{(6)}$ forms an isocahedron (Fig.1(f)) and $\Phi^{(10)}$ forms a dodacahedron (Fig.1(g)). 

\noindent The states found in (II) and (III) are the Platonic solids. All the states in (I) to (III) are the so-called inert states as the distances between Majorana points in these states are independent of interactions. All these states are found in a region containing the parameter set in Table \ref{tab:phasediagram}.  There are also non-inert state in other regions of parameter space.

\subsection{Mean field phase boundary} 
Since there are  many scattering parameters $\{ a_F, F=0,2,\dots, 2f\}$ for large spin systems, it is laborious to explore every corner in the phase diagram numerically. 
However, considerable  insight can be gained by exploring the phase transition boundary using Ginzburg-Landau theory.  

Near the phase boundary, the spin order $\tilde{\rho}_{m_1m_2}$ defined in Eqn.(\ref{nontrivialorder}) is small . We can then expand the free energy in mean field approximation
\begin{equation}
\Omega = - \frac{1}{\beta} \ln \left( \mbox{Tr} e^{- \beta H_{\mbox{\tiny MF}}}\right) - B
\end{equation}
around $\tilde{\rho}_{m_1m_2} = 0$, where $\beta = 1/k_BT$ is the inverse temperature. Here $H_{\mbox{\scriptsize MF}} = H_{\mbox{\scriptsize MF}}^{(0)} + H_{\mbox{\scriptsize MF}}^{(1)}$, with
\begin{eqnarray}
H_{\mbox{\scriptsize MF}}^{(0)} &=& \sum\limits_{\mathbf{k}m} (\epsilon_\mathbf{k} - \mu) c_{\mathbf{k}m}^\dagger c_{\mathbf{k}m},
\\
H_{\mbox{\scriptsize MF}}^{(1)} &=& \sum\limits_{\mathbf{k} m_1m_2}  \tilde{\Gamma}_{m_1m_2} c_{\mathbf{k} m_1}^\dagger c_{\mathbf{k} m_2},
 \\ \label{tildegamma}
 \tilde{\Gamma}_{m_1m_2} &=& 4 \sum\limits_{m_3m_4} \gamma_{m_1m_3m_2m_4}\tilde{\rho}_{m_4m_3},
\end{eqnarray}
and we have restored the condensate energy $B = \langle H- H_{\mbox{\scriptsize MF}}\rangle_{MF}$ to the free energy,
\begin{eqnarray}\nonumber
B &=&
2V\sum_{m_1\dots m_4} \gamma_{m_1m_2m_3m_4} \rho_{m_3m_1}\rho_{m_4m_2}
\\ \label{expandb}
&=& 2V\sum_{m_1 m_2}\tilde{\Gamma}_{m_1m_2}
\tilde{\rho}_{m_2m_1}  + 
Vn^2
\sum_{F=0,2,\dots}^{2f-1} g_F\frac{2F+1}{(2f+1)^2} .
\end{eqnarray}
Terms linear in $\tilde{\rho}_{mm'}$ vanishes due to the identity
\begin{equation}
\langle fm_1; fm_2|FM_F\rangle = (-1)^{f+m_2} \sqrt{\frac{2F+1}{2f+1}} \langle f(-m_2);FM_F|fm_1\rangle
\end{equation}
and completeness relation $1 = \sum_m |fm\rangle \langle fm|$. Similarly, one can show that $\tilde{\Gamma}_{m_1m_2}$ is traceless and Hermitian.

Using the technique of linked cluster expansion \cite{agd}, we have
\begin{eqnarray}\label{fexp}
\Omega &=& \Omega_0 - B - \frac{1}{\beta} \sum\limits_{l=1}^\infty M_l, \quad\quad
\Omega_0 = -\frac{1}{\beta}\ln \mbox{Tr} e^{-\beta H_{\mbox{\tiny MF}}^{(0)}},  \\
\label{mexp}
M_l &=& \frac{(-1)^l}{l!} \int_0^\beta d\tau_1 \dots \int_0^\beta d\tau_l 
\langle {\cal T}_\tau \, H_{\mbox{\scriptsize MF}}^{(1)}(\tau_1) \dots H_{\mbox{\scriptsize MF}}^{(1)}(\tau_l)\rangle_{{c}},\qquad
\end{eqnarray}
where $H_{\mbox{\scriptsize MF}}^{(1)}(\tau) = e^{\tau H_0} V e^{-\tau H_0}$, and ``$\langle\dots\rangle_c$'' means connected diagrams, ${\cal T_\tau}$ is the imaginary-time ordering. Evaluating (\ref{mexp}) using Wick's theorem and keeping up to second order in $\tilde{\rho}_{m_1m_2}$, we have
\begin{equation}\label{omega2}
\Delta\Omega =  -\frac{V }{2}  \left(
 \mbox{Tr} \tilde{\Gamma} \tilde{\rho}
+ \chi(T,\mu)
 \mbox{Tr} \tilde{\Gamma}^2\right),
\end{equation}
where we have treated $\tilde{\Gamma}_{m_1m_2}$ and $\tilde{\rho}_{m_1m_2}$ as matrices.
The susceptibility function is
\begin{equation}
\chi (T,\mu) = \frac{1}{T} \sum_{\bf k} f_{\bf k}(1-f_{\bf k}) =  \int_0^\infty {\rm d}\varepsilon\, D(\varepsilon) \left( -\frac{\partial f(\varepsilon)}{\partial \varepsilon} \right),
\label{chi} \end{equation} 
where  $D(\varepsilon) = 3n\sqrt{\varepsilon}/2\varepsilon_n^{3/2}$ is the density of states, $\varepsilon_n = \hbar^2 k_n^2/2\tilde{m}$, and $f_\mathbf{k} = (e^{(\varepsilon_\mathbf{k}-\mu)/T}+1)^{-1}$ is the Fermi distribution function.  $\chi (T,\mu)$ is always positive and increases as temperature is lowered.  

Now we express (\ref{omega2}) in terms of $\Phi^{(L)}_M$ to see the emergence of each $L$-sector of the spin order.
Note that $\tilde{\rho}_{m_1m_2} $ has the same  expansion  as those of $\rho_{m_1m_2}$ in equation (\ref{rho-phi}), except for the absence of $L=0$ term. Combined with (\ref{tildegamma}), we have
\begin{eqnarray}\label{Gpartial}
\tilde{\Gamma}_{m_1m_2} &=& -\sum_{L=1}^{2f}\sum\limits_{M=-L}^L \Phi^{(L)}_{M} G_L \left( {\cal Y}^{(L)}_M\right)_{m_1m_2},\\
\label{DeltaL}
G_L &=& 2\sum\limits_{F=0,2,\dots}^{2f-1} g_F(2F+1) W(FL).
\end{eqnarray} 
Here we used the identity 
\begin{eqnarray}\nonumber
&&\sum\limits_{m_1m_2 M_F} \langle FM_F|fm_1;f m_a\rangle \langle fm_1| LM;fm_2\rangle \langle fm_2;fm_b|FM_F\rangle  \\
&& =
\langle sm_b|LM;sm_a \rangle (-1)^{2f-F} (2F+1) W(FL),
\end{eqnarray} 
which is derived from the definition of Racah coefficients $ W(ffff;FL) \equiv W(FL)$ \cite{CG}.
Feeding the expansions into (\ref{omega2}), we reach the concise form
\begin{equation}\label{GL2}
\Delta\Omega =  \frac{n^2V}{2} \sum\limits_{L=1}^{2f}\sum\limits_{M=-L}^L \left|\Phi^{(L)}_{M}\right|^2  
  G_{L}\left[1-\chi(T,\mu)G_{L} \right] . 
\end{equation}

Equation(\ref{GL2}) shows that $\Phi^{(L)}$  will emerge if  
\begin{equation}\label{conditionsG}
{\rm (i)} \,\,\,\,\,
G_{L}>0, \,\,\,\,\,\, {\rm and} \,\,\,\,\,  {\rm (ii)} \,\,\,\,\, \chi G_{L}\ge 1, 
\end{equation}
where the equal sign gives the phase boundary.  Up to the quadratic order in $\Phi^{(L)}_M$, all $M$ components are degenerate. 
Higher order terms in $\Phi^{(L)}$ will lift the degeneracy and mix different $L$ components. 
Condition (i) is necessary for the ordered phase to be stable. Consider a single $\Phi^{(L)}_{M}$, which means
$\tilde{\Gamma} = -G_{L} \Phi^{(L)}_{M}{\cal Y}^{(L)}_{M}$.
Then for the ordering to be stable, the  energy $B$ in Eqn.(\ref{expandb}) must be lowered due to the presence of spin orders $\Phi^{(L)}_M$. Since  $B = -2V(\Phi^{(L)}_M{\cal Y}^{(L)}_M)^2 G_L + constant$, the case $G_L>0$ will  ensure the spin ordered phase is  energetically favored over the normal phase. Condition (ii) simply means that the spin-ordered state is at least a local minimum in the free energy functional.

 To help further understanding the conditions (\ref{conditionsG}),  consider a spin-1/2 systems. Here the only non-trivial spin order is $L=1$ ferromagnetic ordering, and $G_1 = g_0$. Then condition (i) reduces to $g_0 >0$, which means the interactions are repulsive. Condition (ii) $\chi g_0 \ge 1$ is the Stoner criterion for ferromagnetic ordering. (Note $\chi\rightarrow D(\varepsilon_F)$ at $T=0$). For higher spin systems, $G_L$ represents the total interaction strength responsible for triggering the spin order $|\Phi^{(L)}\rangle$. Since the Racah coefficients $W(FL)$ can be both positive and negative, not all $g_F$'s have to be repulsive 
 in order to create a critical total interaction $G_L$ to start the spin order, unlike the spin-1/2 case.

Experimental parameters are usually expressed by scattering length $a_F$. Thus, we define the total scattering length $A_L$ 
\begin{equation}
G_{L} = \frac{8 \pi \hbar^2 A_{L}}{\tilde{m}}, \,\,\,\,\,\, A_{L} = \sum_{F} a^{}_{F} (2F+1) W(F,L). 
\label{baraL} \end{equation}
Next, we note that $\chi(T,\mu)$ in Eqn.(\ref{chi})   
has the dimension of density of state. It scales as $\sqrt{\mu}$ and hence represents a momentum scale. We can then define a wavevector $k(T, \mu)$ as  $\chi(T, \mu)  = (\tilde{m}/2 \pi^2 \hbar^2)k(T,\mu)$. (Note that $k(0,\mu) = k_{n}$). Condition (ii) then becomes
\begin{equation}
k(T, \mu) A_{L} \geq  \frac{\pi}{4}, \,\,\,\,\, {\rm or} \,\,\,\,\, k_{n}A_{L} \geq  \frac{\pi}{4}\frac{k_{n}}{k(T, \mu)}, 
\label{Tc} \end{equation}
where the equal sign gives the phase boundary. 

\begin{figure}[hbtp]
\includegraphics[height=5cm,width=7cm]{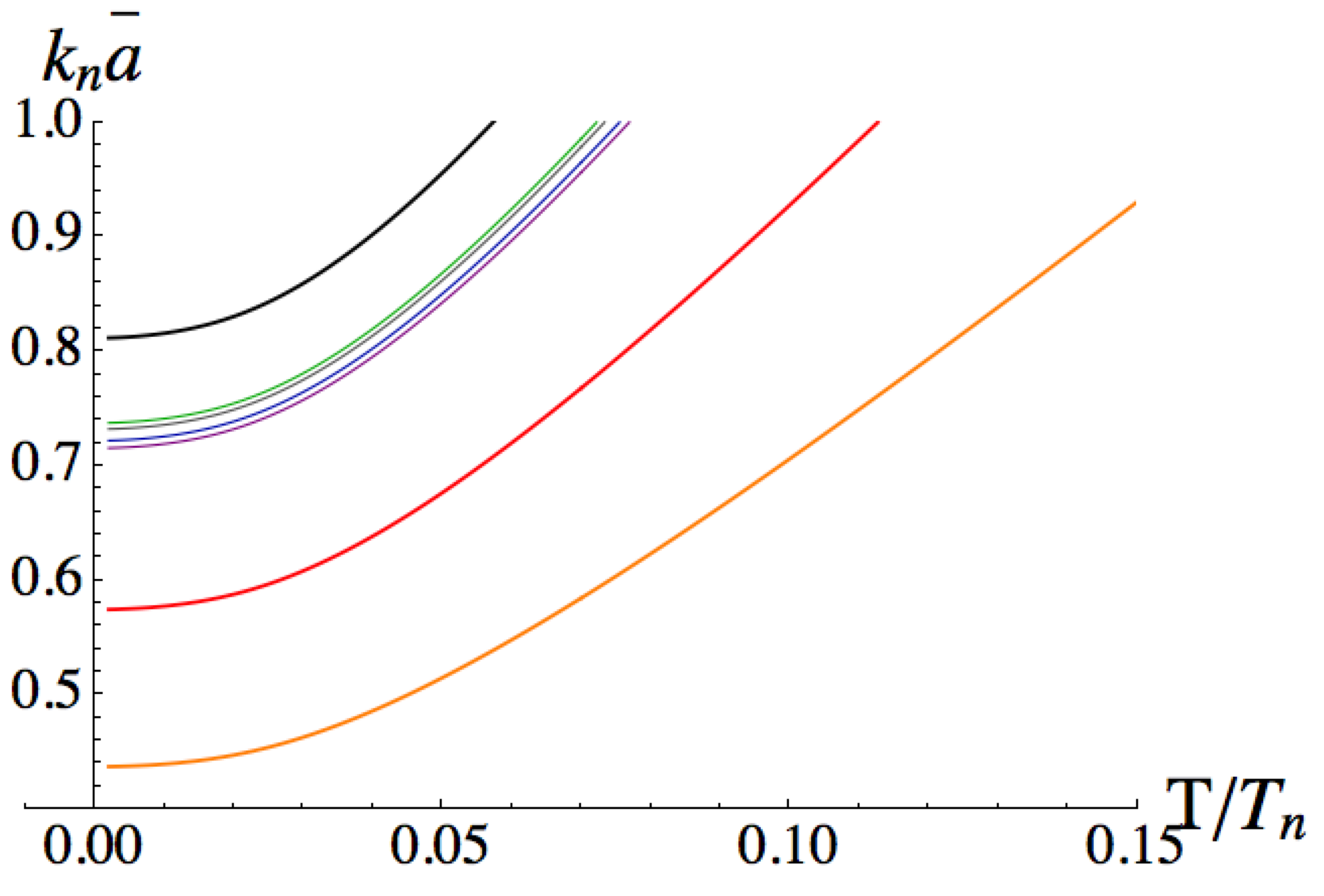}
\caption{   The transition temperature for spin order $|\Phi^{(L)}\rangle$ for $L=1,2,6,3,5,4,7$, (from bottom to top), for a spin $f=21/2$ Fermi gas. } \label{fig1}
\end{figure}

Certainly, the larger the $A_{L}$, the easier for the $L$-th spin order to emerge. However, for small gas parameters $k_{n}a_{F}<1$, it is not clear whether Eq.(\ref{Tc}) can be satisfied.  On the other hand, one sees from equation (\ref{baraL}) that  $A_{L}$ will be maximized if the sign of $a_{F}$ matches that of the Racah coefficient $W(F,L)$.  To demonstrate this effect, we consider a set of  $a_{F}$'s with the same magnitude $\bar{a}$ with a sign matching that of  $W(F, L)$. 
equation (\ref{Tc})
then becomes 
\begin{equation}
k_{n} \bar{a} \geq  \frac{\pi/4}{  \sum_{F=0,2, .}^{2f}  (2F+1) |W(F,L)|}\frac{k_{n}}{k(T, \mu)}.  
\label{newTc} \end{equation}
This condition is plotted in Figure 2 for a spin $f=21/2$ Fermi gas. It shows  spin orders as high as $L=7$ can emerge at the phase boundary for   $k^{}_{n} \bar{a} <1$.  While equation (\ref{newTc}) is sufficient for the appearance of $\Phi^{(L)}$, it is not necessary. Once a low order $\Phi^{(L)}$ is present, say, $L=1$, 
higher $L$ spin order can emerge through non-linear coupling as temperature is lowered. Finally, we note from Eqn.(\ref{newTc}) that the larger  the spin $f$ of the fermions, the larger the sum in Eqn.(\ref{newTc}), and the smaller the gas parameter $k_{n} \bar{a}$ needed to activate the spin order.

\section{Experimental Determination of the density matrix} 
Since the diagonal element of the density matrix is the spin population along a specified spin quantization axis, say $\hat{\bf z}$, they can be  determined by the Stern-Gerlach method. To access the off diagonal elements, one can apply a magnetic pulse to rotate $\rho$ to $\rho' = D\rho D^{\dagger}$, where $D$ is a rotational matrix, see Eqn.(\ref{rho-trans}).  The diagonal elements of $\rho'$ will then contain information of the off-diagonal elements of $\rho$ due to the rotation $D$.  By repeating the measurement of diagonal matrix elements for different $D$'s, one can then extract the information of the off-diagonal matrix elements of the original density matrix $\rho$.

\section{Concluding Remarks}
Large spin quantum gases are fertile grounds for new quantum matter. Here, we have pointed out the very rich spin order possible in large spin fermions, most of which have no analog in electron matters. We show that the spin order in different sectors can be conveniently described as Majorana antipodal points on a sequence of spheres representing the spin order of different particle-hole angular momenta. Our model calculations show that some of these orders can take the form of Platonic solids, which are structures that exist within certain {\em region} of the parameter space instead of a single point. These structures are therefore robust and will have good chance to be realized.

\begin{center}
{ \bf ACKNOWLEDGEMENTS}
\end{center}
This work is supported by DARPA under the Army Research Office Grant Nos. W911NF-07-1-0464, W911NF0710576, and NSF Grant DMR-1309615.

\end{document}